\documentclass[12pt]{article}
\pdfoutput=1
\usepackage{amsmath}
\usepackage{amssymb}
\usepackage{graphicx}
\usepackage{subcaption}
\usepackage{url}
\usepackage{arydshln}
\usepackage[sort&compress, numbers, merge]{natbib}

\newcommand{\be}{\begin{equation}}
\newcommand{\ee}{\end{equation}}

\newcommand{\bea}{\begin{eqnarray}}
\newcommand{\eea}{\end{eqnarray}}
\newcommand{\mat}{\begin{pmatrix}}
\newcommand{\rix}{\end{pmatrix}}

\renewcommand{\bar}{\overline}

\def\beq{\begin{equation}}
\def\eeq{\end{equation}}
\def\beqn{\begin{equation}}
\def\eeqn{\end{equation}}

\usepackage[colorlinks=true, linkcolor=magenta, citecolor=blue,
urlcolor=black]{hyperref}

\begin{document}

\begin{titlepage}
\begin{center}

\vspace{3.0cm}
{\large \bf Final state Sommerfeld effect on dark matter relic abundance}

\vspace{1.0cm}
{\bf Xiaoyi Cui}$^{(a)}$ and 
{\bf Feng Luo}$^{(b)}$ 
\vspace{1.0cm}

{\it
$^{(a)}${School of Mathematics (Zhuhai), Sun Yat-sen University, Zhuhai 519082, China}\\
$^{(b)}${School of Physics and Astronomy, Sun Yat-sen University, Zhuhai 519082, China}\\
}

\vspace{1cm}
\abstract{ 
If the annihilation products of dark matter (DM) are non-relativistic and if there is some long-range force between them, there can be Sommerfeld effect for the final state particles.
We study this effect on DM relic abundance in the thermal freeze-out scenario. 
As a proof of concept, we consider the case of a DM pair annihilation into a final state pair, assuming that the mutual interactions between the two final state particles give rise to a Coulomb-like potential, and that the masses of the initial and final state particles are similar, so that both the initial and final state particles are non-relativistic.     
The size of the final state Sommerfeld (FSS) effect depends on the strength of the potential, as well as on the mass ratio of the final and initial state particles.
We find that the impact of the FSS effect on DM relic abundance can be significant, and an electroweak sized long-range interaction is large enough to make a correction well beyond the observational accuracy.
Another feature of the FSS effect is that it could be suppressed when its time scale is longer than the lifetime of the final state particles. As a corollary, we also study in the DM coannihilation scenario where the initial state Sommerfeld effect between two coannihilators could be reduced due to their instability, which may need to be taken into account for an accurate calculation of the DM relic abundance.  
}
\end{center}

\end{titlepage}

\setcounter{footnote}{0}

\section{Introduction}
The relic abundance of Weakly Interacting Massive Particle (WIMP) dark matter (DM) is usually given by the thermal freeze-out mechanism~\cite{Lee:1977ua, Kolb:1990vq}.
During freeze-out, the annihilating DM particles are moving with non-relativistic velocities.
If there is some long-range interaction between two annihilating particles, such that the two-body wave function is modified from the plane wave, the annihilation cross sections and the DM relic abundance are affected. 
This non-relativistic quantum mechanics effect --- the so-called Sommerfeld effect~\cite{sommerfeld1931} --- on DM relic abundance has been well-studied.
Also, in coannihilation scenarios~\cite{Griest:1990kh} the long-range interaction strengths between coannihilators can often be larger than the ones between DM particles; therefore, the Sommerfeld effect can be significant in coannihilation scenarios (see e.g.~\cite{hep-ph/9806361, hep-ph/0610249, 0706.4071, 1005.4678, 1010.2172, deSimone:2014pda, 1403.0715, 1404.5571, 1503.07142, 1601.04718, 1611.08133, 1812.02066}). 

Previous works about the Sommerfeld effect have been focusing on the initial DM or coannihilator pairs.
However, the two conditions of the Sommerfeld effect, namely, non-relativistic motion and long-range interaction, may also apply to the annihilation products. 
In this work, we study the Sommerfeld effect of the final state particles on DM relic abundance. 

First, it is possible that the annihilation products can move with non-relativistic velocities. 
An example is that the annihilation products have masses close to the DM or coannihilators in a 2-to-2 reaction. 
The annihilation products can be either the Standard Model (SM) particles or new particles in beyond the Standard Model (BSM) theories. 
The mass degeneracy between the initial and final state particles can be either a coincidence or from a symmetry. 
For example, if supersymmetry were not broken or only slightly broken, the $R$-parity odd sparticles would have been nearly mass-degenerate with their corresponding $R$-parity even partners which can be the annihilation products of the former.
In particular, at finite temperature during thermal freeze-out, the initial annihilating particles in the tail of the Maxwell-Boltzmann velocity distribution can have enough energy to annihilate into heavier particles. 
As one of the ``three exceptions"~\cite{Griest:1990kh}, annihilation into forbidden channels is important when these channels dominate the total annihilation cross section and determine the DM relic abundance. 
This scenario has been further developed recently under the names of ``Forbidden DM"~\cite{1505.07107} and ``Impeded DM"~\cite{1609.02147}. 

Second, it is natural that there is some long-range interaction between the annihilation products. A typical situation is that the final state is a particle-antiparticle pair, and gauge bosons can be exchanged between them. The gauge groups could be either the ones in the SM or the ones in BSM theories. 
Exchanges of the Higgs or some other new scalars are also possible. 

We will show that for non-relativistically moving annihilation products with long-range interaction strengths comparable to the SM ones, the final state Sommerfeld (FSS) factors are large enough to affect significantly the DM relic abundance, well beyond the percent level accuracy of the observationally determined value~\cite{Aghanim:2018eyx}. 

In fact, the FSS factor has been discussed routinely in collider productions of charged particles near the production thresholds~\footnote{The Sommerfeld effect is often called the Coulomb effect in this context.}, including $W^+W^-$, $t \bar{t}$, charged leptons, baryons and heavy mesons (see e.g.~\cite{hep-ph/9501214, hep-ph/9904278, hep-ph/0301076, 0711.1725, 1111.4308, 1303.0653}), where the long-range interactions are induced by the exchanges of the SM gauge bosons. 

There is another feature that needs to be considered in the calculations of the FSS effect. 
While the DM in the initial state is stable, the final state particles may decay. 
The FSS effect is ineffective if its time scale is longer than the lifetimes of the final state particles. 
This time scale is given by the inverse of the kinetic energy of the final state particles~\cite{Fadin:1993kg, Bardin:1993mc, Fadin:1993kt, hep-ph/9501214}. 
Consequently, for the case that the final state is a particle-antiparticle pair or are two same particles, with mass $m_2$ and decay width $\Gamma_2$, one may introduce a cut-off velocity $\sim\sqrt{\Gamma_2/m_2}$, below which the FSS effect is suppressed.

In fact, a similar velocity cut-off also applies to the initial state Sommerfeld (ISS) effect between two coannihilators, which can convert to DM particles through scatterings and decays. 
Suppose that there is a direct coupling among a DM, a coannihilator and a massless particle, at the leading order the scattering rate takes the form of $\sim c_{\rm scatt} T^3/m_1^2$, while the decay rate takes the form of $\sim c_{\rm decay} \Delta m$, where $m_1$ is the DM particle mass, $T$ is the temperature, $\Delta m$ is the mass difference between the coannihilator and the DM particle, and the massless parameters $c_{\rm scatt}$ and $c_{\rm decay}$ depend on the couplings and mixings in a specific BSM model. 
While the time scale of scatterings is usually larger than that of the ISS effect, we find that for not very small $\Delta m$ the coannihilator decay time can be smaller than the latter and the ISS effect can be suppressed.
Since $\Delta m$ determines the signatures of collider search for WIMP, the modification of DM relic abundance due to the suppression of the coannihilators' Sommerfeld effect may need to be taken into account, when one is looking for viable and interesting parameter regions in BSM models. 

The rest of our paper is organized as follows. 
In Section~\ref{sec:fss}, we calculate the thermally averaged FSS factor in WIMP DM annihilation during thermal freeze-out, estimate its impact on DM relic abundance, and discuss the modification induced by decays of the final state particles. 
In Section~\ref{sec:iss}, we estimate in DM coannihilation scenarios the effect of the coannihilators' instability on the ISS factor between two coannihilators. 
We summarize our conclusions in Section~\ref{sec:sum}.
 
\section{Final state Sommerfeld effect}
\label{sec:fss}
The origin of the FSS effect is the same as the more familiar initial state one, namely, the plane wave functions are modified due to some long-range force between non-relativistically moving final state particles. 
To illustrate the physics, in this paper we consider a simple Coulomb-like potential between two final state particles in a 2-to-2 annihilation, and we focus on the $s$-wave Sommerfeld effect. 
We assume that the two incoming particles have the same mass, $m_1$, and the two outgoing particles have the same mass, $m_2$. 

\subsection{Thermally averaged FSS factor}
\label{subsec: FSS factor}
Following the notations in~\cite{Griest:1990kh}, the product of the annihilation cross section and the relative velocity of the incoming particles in the center-of-mass frame is written as $(\sigma v)_{\rm w/o\, FSS} = a^\prime v_2$, where $a^\prime$ is a constant and $v_2$ is the velocity of one of the final state particles. 
The subscript `w/o FSS' indicates that the FSS factor is not included. 
The dependence of $v_2$ comes from doing the phase space integration of the two outgoing particles. 
$v_2$ relates with $v$, $m_1$ and $m_2$ through
\beq
v_2 = \sqrt{1-z^2 + z^2 v^2/4} \, ,
\label{eq: v2}
\eeq
where $z \equiv m_2/m_1$.
For $z \le 1$, the reaction is kinematically allowed for any value of $v$, and the minimum value of $v_2$ is $\sqrt{1-z^2}$. 
For $z > 1$, the reaction is kinematically allowed only for $v \ge 2 \sqrt{1-1/z^2}$, and the minimum value of $v_2$ is $0$. 
Therefore, when $m_2$ is not too much smaller than $m_1$, say, $z \gtrsim 0.8$, the final state particles can move with non-relativistic velocities, considering that the incoming particles move non-relativistically during the freeze-out and thereafter. 

A Coulomb-like potential between the two non-relativistically moving final state particles, $V(r) = - \alpha_f / r$, modifies the otherwise two-body plane wave function and leads to the FSS effect. 
Depending on the charges or quantum states~\footnote{For example, for the color SU(3) interaction, the value of $\alpha_f$ is determined by the quadratic Casimir coefficients of the color representations of both the individual particles and the two as a single color state~\cite{1611.08133}.} of the final state particles under this long-range interaction, $\alpha_f$ can be either positive or negative. 
The Sommerfeld-corrected $s$-wave cross section takes the form of $(\sigma v)_{\rm with \, FSS} = a^\prime S_f v_2$, where the FSS factor is
\beq
S_f = \frac{\pi \alpha_f / v_2}{1 - e^{-\pi \alpha_f / v_2}} \, . 
\label{eq: S_f}
\eeq
The thermally averaged cross section is
\beq
\langle \sigma v \rangle_{\rm with \, FSS} 
=  a^\prime \langle S_f v_2 \rangle 
= a^\prime \int_{v_{\rm min}}^{\infty} f(v) S_f v_2 \, dv \, ,
\label{eq: thermal S_f v_2}
\eeq
where $v_{\rm min} = 0$ for $z \le 1$, and $v_{\rm min} = 2 \sqrt{1-1/z^2}$ for $z > 1$. 
$f(v)$ is the Maxwell-Boltzmann distribution for the relative velocity $v$, given as
\beq
f(v) = \left(\frac{m_1/2}{2 \pi T}\right)^{3/2} 4 \pi v^2 e^{- \frac{v^2 m_1/2}{2 T}} = \frac{x^{3/2}}{2 \sqrt{\pi}} v^2 e^{- x v^2/4} \, ,
\label{eq: MB}
\eeq
where $x \equiv m_1/T$. 
To facilitate the integration, we do a change of variables, $t \equiv (v^2 - v_{\rm min}^2) x / 4$, so that $v_2 = \sqrt{1-z^2 + z^2 \left(t/x + v_{\rm min}^2 / 4 \right)}$, and the $\langle S_f v_2 \rangle$ in Eq.~(\ref{eq: thermal S_f v_2}) can be written as
\beq
\langle S_f v_2 \rangle = 2 \sqrt{x / \pi} e^{- v_{\rm min}^2 x / 4} \int_0^\infty \left( \frac{t}{x} + \frac{v_{\rm min}^2}{4}\right)^{1/2} e^{-t} S_f v_2 \, dt \, .
\label{eq: thermal S_f v_2 int_t}
\eeq 
Since $S_f$ is applicable to a non-relativistic $v_2$, we choose to turn off the FSS effect for $v_2 > 0.6$. 
That is, when using Eq.~(\ref{eq: thermal S_f v_2 int_t}) in the following calculations, we substitute $S_f$ by $\left[(S_f - 1) H(0.6 - v_2) + 1\right]$, where $H(0.6 - v_2)$ is the Heaviside step function. 
This choice means that we only consider the FSS effect for $z \ge 0.8$. 

If not considering the FSS effect for all $v_2$, i.e., in the $\alpha_f \to 0$ limit, the thermally averaged cross section becomes 
\beq
\langle \sigma v \rangle_{\rm w/o\, FSS} 
= a^\prime \langle v_2 \rangle
= a^\prime \int_{v_{\rm min}}^{\infty} f(v) v_2 \, dv 
= a^\prime 2 \sqrt{x / \pi} e^{- v_{\rm min}^2 x / 4} \int_0^\infty \left( \frac{t}{x} + \frac{v_{\rm min}^2}{4}\right)^{1/2} e^{-t} v_2 \, dt 
\, ,
\label{eq: thermal v_2}
\eeq
for which the integration can be done analytically, as shown in Eqs.~(24) and (25) of~\cite{Griest:1990kh}.  

\begin{figure}
\begin{center}
\begin{tabular}{c c}
\hspace{-0.3cm}
\includegraphics[height=5.8cm]{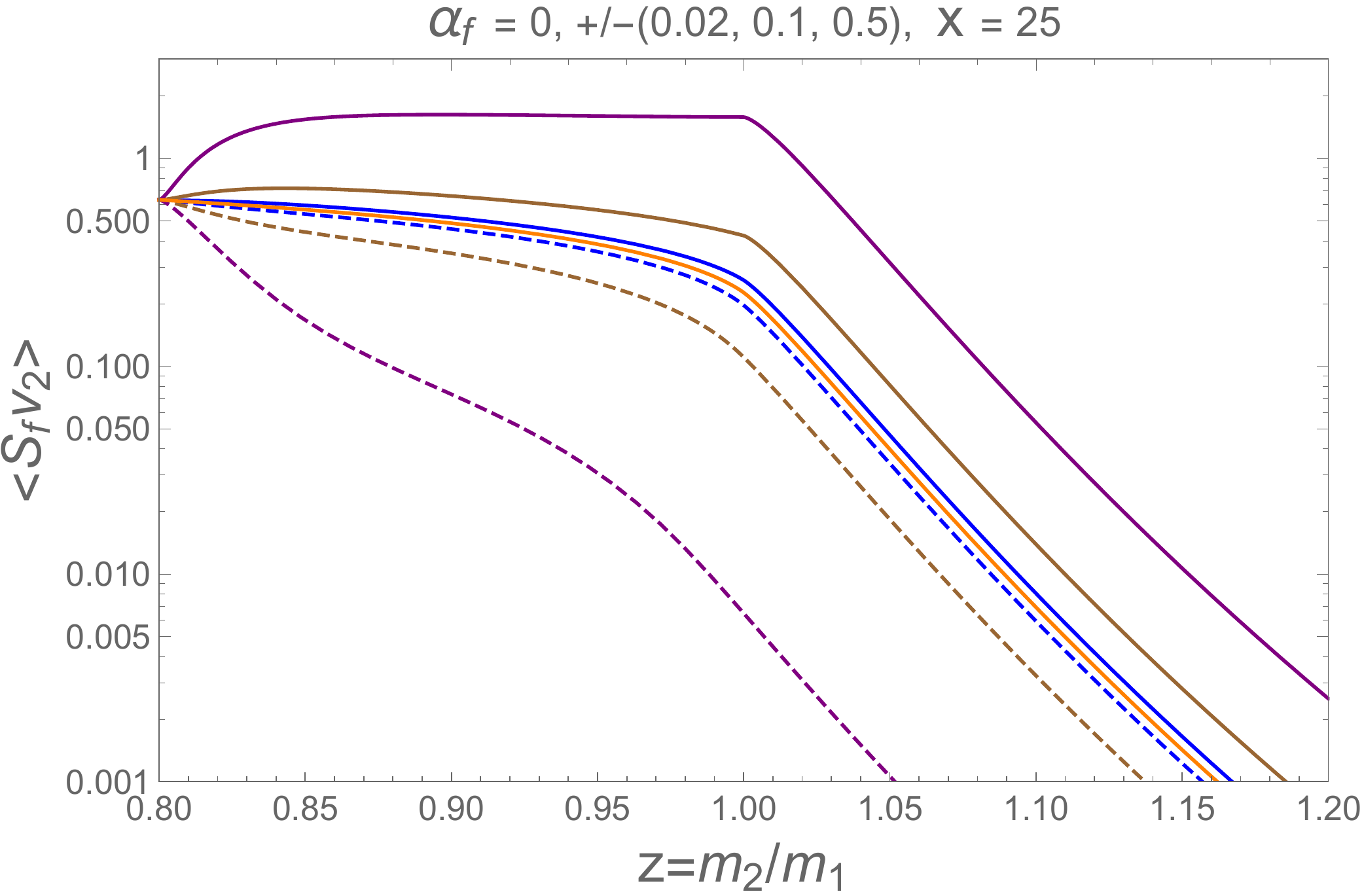} & 
\hspace{-0.3cm}
\includegraphics[height=5.8cm]{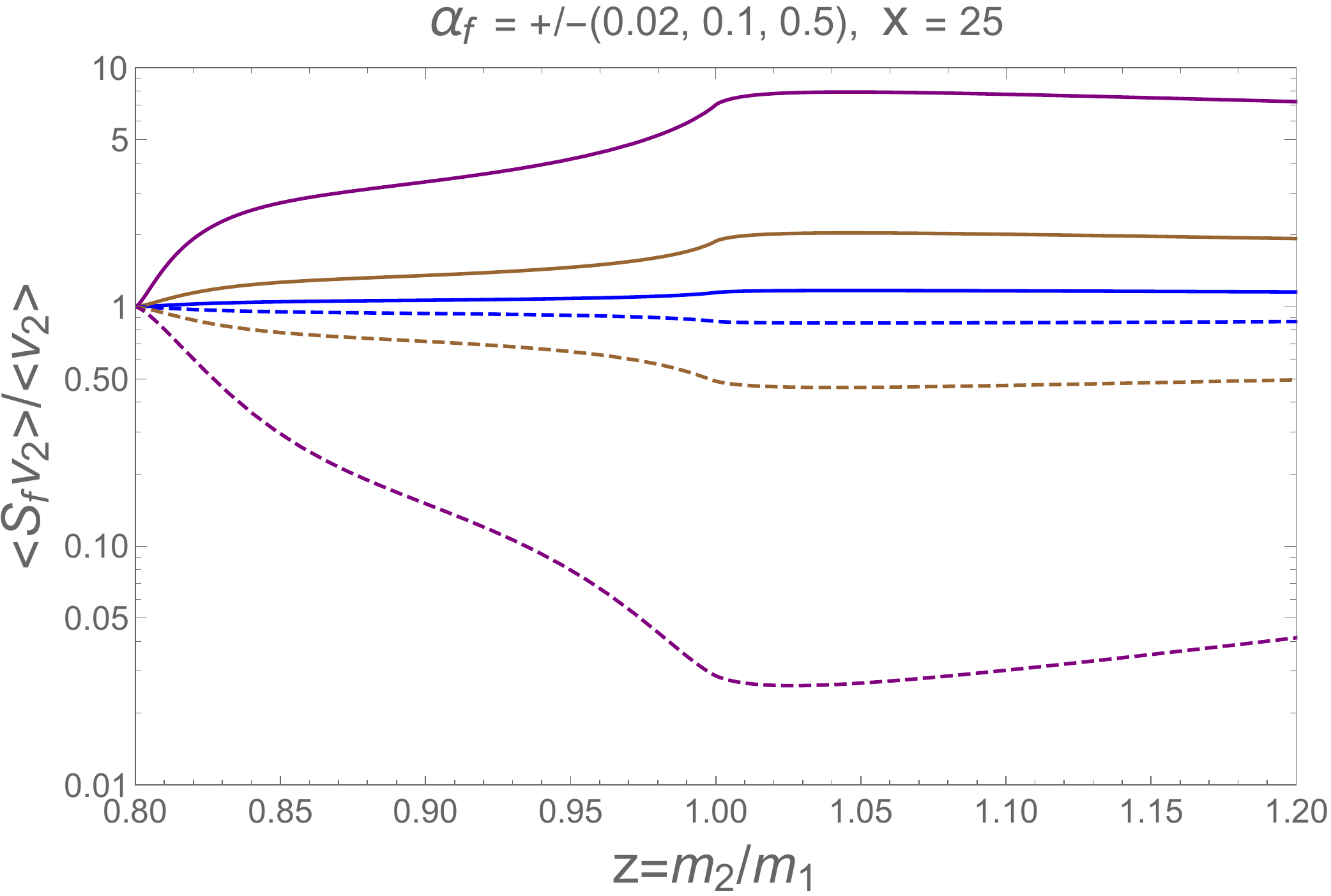} \\
\end{tabular}
\end{center}   
\caption{\label{fig: sigmav_vs_z}\it
Left panel: the thermally averaged FSS-corrected $s$-wave cross section (up to the factor $a^\prime$), as a function of the mass ratio of the final and initial state particles, for a typical freeze-out value $x = m_1/T = 25$. 
The solid blue, brown and purple lines are for $\alpha_f = + \, 0.02$, $+ \, 0.1$ and $+ \, 0.5$, respectively. 
The dashed blue, brown and purple lines are for $\alpha_f = - \, 0.02$, $- \, 0.1$ and $- \, 0.5$, respectively. 
The orange line in the middle is for $\alpha_f = 0$, namely, the $\langle v_2 \rangle$ curve. 
Right panel: the ratio of the thermally averaged cross sections with and without the FSS effect. 
That is, it is the same as the left panel, but with the blue, brown and purple lines normalized to the orange line. 
}
\end{figure}

The left panel of Fig.~\ref{fig: sigmav_vs_z} plots $\langle S_f v_2 \rangle$ as a function of $z$ at a typical freeze-out value of $x=25$, for several choices of $\alpha_f$. 
The solid blue, brown and purple lines are for $\alpha_f = + \, 0.02$, $+ \, 0.1$ and $+ \, 0.5$, respectively. 
The dashed blue, brown and purple lines are for $\alpha_f = - \, 0.02$, $- \, 0.1$ and $- \, 0.5$, respectively. 
A positive (negative) $\alpha_f$ results in a Sommerfeld enhancement (suppression) for the thermally averaged cross section. 
The orange line in the middle is for $\alpha_f = 0$, that is, $\langle v_2 \rangle$; it is the same as the solid line in the upper panel of Fig.~2 in~\cite{Griest:1990kh}. 
All lines merge at $z = 0.8$, since below this value the minimum $v_2$ is larger than 0.6 and we choose to turn off the FSS effect, as mentioned above. 
During the freeze-out and thereafter, typical incoming particles are non-relativistic, so that $v_2$ decreases with the increase of $z$, according to Eq.~(\ref{eq: v2}). 
This explains the general trend of the curves. 

To see the Sommerfeld effect more clearly, we show in the right panel of Fig.~\ref{fig: sigmav_vs_z} the ratio of $\langle S_f v_2 \rangle$ to $\langle v_2 \rangle$, as a function of $z$. 
That is, at each $z$, the values on the solid and dashed blue, brown and purple lines are the corresponding ones in the left panel divided by the value on the orange line. 
Starting from $z = 0.8$, for a given $\alpha_f$ the FSS effect increases with the increase of $z$ until the latter is slightly bigger than $1$, and after that it mildly decreases. 
This behavior can be understood from the fact that for $z > 1$ the fraction of the incoming particles that have enough kinetic energy to activate the reaction becomes smaller for larger $z$, and that the deviation of the FSS factor from $1$ is larger for smaller $v_2$ according to Eq.~(\ref{eq: S_f}), together with the above-mentioned fact that $v_2$ decreases with the increase of $z$. 
At $x = 25$ and for $z = 1$, there is a $\sim 15\%$ Sommerfeld enhancement for the thermally averaged cross section for an electroweak interaction sized $\alpha_f = + \, 0.02$, a factor of $\sim 2$ enhancement for a strong interaction sized $\alpha_f = + \, 0.1$, and a factor of $\sim 7$ enhancement for an even stronger interaction $\alpha_f = + \, 0.5$; while for $\alpha_f = - \, 0.02, - \,0.1$ and $- \, 0.5$, the suppression factor is about $0.87$, $0.5$ and $0.03$, respectively. 
 
\subsection{FSS effect on DM relic abundance}
\label{subsec: DM relic abundance}
Fig.~\ref{fig: sigmav_vs_z} gives us a hint that the FSS effect may bring a noticeable correction to the DM relic abundance. 
After the freeze-out, the DM density can continue decreasing by a factor of a few or even an order of magnitude and more. Therefore, we need to study the Boltzmann equation which describes the evolution of the DM density. 
By introducing the {\it yield}, which is the ratio of the DM number density to the entropy density, $Y_1 \equiv n_1/s$, 
the Boltzmann equation can be written as
\bea
\frac{dY_1}{dx}=-\frac{xs}{H(m_1)}\left(1+\frac{T}{3g_{\ast s}}\frac{dg_{\ast s}}{dT}\right)\langle \sigma  v\rangle\left(Y_1^2- Y^2_{1, eq}\right) \, ,
\label{eq: single_Boltzmann}
\eea
where
\bea
s = {2 \pi^2 \over 45} g_{\ast s} T^3 = {2 \pi^2 \over 45} g_{\ast s} m_1^3 / x^3 \, , \; \;\; \;\; \;  H(m_1) \equiv H(T) x^2 = \left({4 \pi^3 G_N g_\ast \over 45}\right)^{1 \over 2} m_1^2 \, .
\label{eq: def_s_and_Hm}
\eea
$G_N$ is the gravitational constant, $H(T)$ is the Hubble parameter, $g_{\ast s}$ and $g_\ast$ are the numbers of effectively massless degrees of freedom associated with the entropy density and the energy density, respectively. 
$Y_{1, eq}$ is the equilibrium value of the yield, $Y_{1, eq} = n_{1, eq}/s$, where 
\be
n_{1, eq} = \frac{T}{2 \pi^2} g_1 m_1^2 K_2 (x) \, .
\label{eq: def_neq}
\ee 
$g_1$ is the DM degrees of freedom, and $K_2 (x)$ is the modified Bessel function of the second kind. At $x \gg 1$, $n_{1, eq} \approx g_1 \left(\frac{m_1 T}{2 \pi}\right)^{3/2} e^{-m_1/T}$. By integrating Eq.~(\ref{eq: single_Boltzmann}) from a small value of $x$ before the freeze-out when $Y_1 = Y_{1, eq}$, to its current value which is essentially $x \to \infty$, we get the yield of today, $Y_{1,0}$. The DM relic abundance is given as~\cite{Srednicki:1988ce, Gondolo:1990dk, Edsjo:1997bg}~\footnote{If the two incoming particles are not identical, for example, they are Dirac fermion and anti-fermion, then in Eq.~(\ref{eq: single_Boltzmann}) $Y_1$ and $Y_{1, eq}$ include both the contributions from particles and antiparticles, assuming that there is no asymmetry between their number densities. Also, a factor of $\frac{1}{2}$ may need to be introduced to $\langle \sigma  v\rangle$, as explained in the Appendix of~\cite{Srednicki:1988ce}. Nevertheless, in this work we are interested in the {\it ratio} of DM relic densities with and without including the Sommerfeld effect, and this factor cancels out.} 

\be
\Omega h^2 = 2.755 \times 10^8 \frac{m_1}{\text {GeV}} Y_{1, 0} \, ,
\label{eq: relic abundance}
\ee
where $h$ is the present-day dimensionless Hubble parameter.

We note that by writing the term $Y_{1, eq}^2$ in Eq.~(\ref{eq: single_Boltzmann}), we have assumed the usual condition that the annihilation products quickly thermalize and that their number densities equal the thermal equilibrium values. This assumption ensures that the information about the final state particles' number density $n_{2}$ and its equilibrium value $n_{2, eq}$ does not appear in the Boltzmann equation, due to the principle of detailed balance, $\langle \sigma v \rangle_\rightarrow n_{1, eq}^2 = \langle \sigma v \rangle_{\leftarrow} n_{2, eq}^2$, where $\langle \sigma v \rangle_{\rightarrow} \equiv \langle \sigma v \rangle$ is the forward reaction and $\langle \sigma v \rangle_{\leftarrow}$ is the backward reaction. 
While this assumption is true for SM final products, one might need to check its validity when the annihilation products are new particles in  BSM models. Without committing to specific BSM models, in this work we make this assumption and study the FSS effect in the simplest situation. 

We are interested in the ratio of DM relic densities with and without including the FSS effect, 
$\Omega / \Omega_{\rm w/o \, S_f}$. 
A good estimation of the ratio can be obtained by using an approximate form of Eq.~(\ref{eq: single_Boltzmann}), 
\bea
\frac{dY_1}{dx} \approx -\frac{xs}{H(m_1)}\left(1+\frac{T}{3g_{\ast s}}\frac{dg_{\ast s}}{dT}\right)\langle \sigma  v\rangle Y_1^2 \, ,
\label{eq: single_Boltzmann_approx}
\eea
which is valid after the freeze-out, since $Y_{1, eq}$ quickly becomes negligible compared to $Y_1$. Eq.~(\ref{eq: single_Boltzmann_approx}) has the solution 
\bea
Y_1(x_2) \approx \left[\frac{1}{Y(x_{\rm fo})}+\int^{x_2}_{x_{\rm fo}} \frac{xs}{H(m_1)}\left(1+\frac{T}{3g_{\ast s}}\frac{dg_{\ast s}}{dT}\right)\langle \sigma v\rangle dx \right]^{-1} \, ,
\label{eq: approx_solution}
\eea
where $x_{\rm fo}$ is the freeze-out value of $x$, defined when $Y_{1} (x_{\rm fo}) - Y_{1, eq} (x_{\rm fo}) \equiv \kappa Y_{1, eq} (x_{\rm fo}) $ for a numerical constant $\kappa$ of order unity. 

Since the final yield $Y_{1,0}$ is usually much smaller than $Y_{1} (x_{\rm fo})$, we can make a further approximation to get  
\bea
Y_{1,0} \approx \left[\int^\infty_{x_{\rm fo}} \frac{xs}{H(m_1)}\left(1+\frac{T}{3g_{\ast s}}\frac{dg_{\ast s}}{dT}\right)\langle \sigma v\rangle dx \right]^{-1} \, .
\label{eq: Y_today_approx}
\eea
Therefore, the ratio we want is 
\bea
\frac{\Omega}{\Omega_{\rm w/o \, S_f}} = \frac{Y_{1,0}}{Y_{1,0, \rm w/o \, S_f}} \approx 
\frac{\int^\infty_{x_{\rm fo,n}} \frac{\langle \sigma v\rangle_{\rm w/o \, FSS}}{x^2} dx} 
{\int^\infty_{x_{\rm fo,y}} \frac{\langle \sigma v\rangle_{\rm with \, FSS}}{x^2} dx} 
= \frac{\int^\infty_{x_{\rm fo,n}} \frac{\langle v_2 \rangle}{x^2} dx} 
{\int^\infty_{x_{\rm fo,y}} \frac{\langle S_f v_2 \rangle}{x^2} dx}   \, ,
\label{eq: ratio}
\eea
in which we have dropped the factor $\left(1+\frac{T}{3g_{\ast s}}\frac{dg_{\ast s}}{dT}\right)$ in the integrals, since $|\frac{T}{3g_{\ast s}}\frac{dg_{\ast s}}{dT}| \ll 1$ except in the narrow range around the quark-hadron phase transition temperature $\sim 200 \, {\rm MeV}$~\cite{Srednicki:1988ce, Saikawa:2018rcs}. 
Unless $m_1 / x_{\rm fo}$ falls into this temperature range, neglecting this factor is of no harm to our estimation of the ratio. 

In Eq.~(\ref{eq: ratio}) the lower limits of the two integrals, $x_{\rm fo,y}$ and $x_{\rm fo,n}$, are $x_{\rm fo}$ with and without the inclusion of the FSS effect. Since $(x_{\rm fo} + \frac{1}{2} \ln x_{\rm fo} - \ln \langle \sigma v\rangle_{x_{\rm fo}})$ is roughly a constant (see e.g. Eq.~(2) of~\cite{Griest:1990kh}), where $\langle \sigma v\rangle_{x_{\rm fo}}$ means the value of $\langle \sigma v\rangle$ evaluated at $x_{\rm fo}$, then 
\bea
x_{\rm fo,y} + \frac{1}{2} \ln x_{\rm fo,y} - \ln\langle S_f v_2 \rangle_{x_{\rm fo,y}} \approx 
x_{\rm fo,n} + \frac{1}{2} \ln x_{\rm fo,n} - \ln\langle v_2 \rangle_{x_{\rm fo,n}} \, . 
\label{eq: x_fo_y,n relation}
\eea
Since $|x_{\rm fo,y} - x_{\rm fo,n}| \ll x_{\rm fo,n}$, up to the first order of $(x_{\rm fo,y} - x_{\rm fo,n})/x_{\rm fo,n}$ we get
\bea
x_{\rm fo,y} \approx x_{\rm fo,n} + \left(1 + \frac{1}{2 x_{\rm fo,n}} - \frac{\langle S_f v_2 \rangle^\prime_{x_{\rm fo,n}}}{\langle S_f v_2 \rangle_{x_{\rm fo,n}}} \right)^{-1} \ln \frac{\langle S_f v_2 \rangle_{x_{\rm fo,n}}}{\langle v_2 \rangle_{x_{\rm fo,n}}}    
\, ,
\label{eq: x_fo,y}
\eea
where $\langle S_f v_2 \rangle^\prime_{x_{\rm fo,n}}$ means the derivative of $\langle S_f v_2 \rangle$ with respect to $x$, evaluated at ${x_{\rm fo,n}}$. Using this relation between $x_{\rm fo,y}$ and $x_{\rm fo,n}$, we see from Eq.~(\ref{eq: ratio}) that an estimation of $\Omega / \Omega_{\rm w/o \, S_f}$ only depends on the values of $x_{\rm fo,n}$, $z$ and $\alpha_f$. 

\begin{figure}
\begin{center}
\begin{tabular}{c c}
\hspace{-0.3cm}
\includegraphics[height=5.8cm]{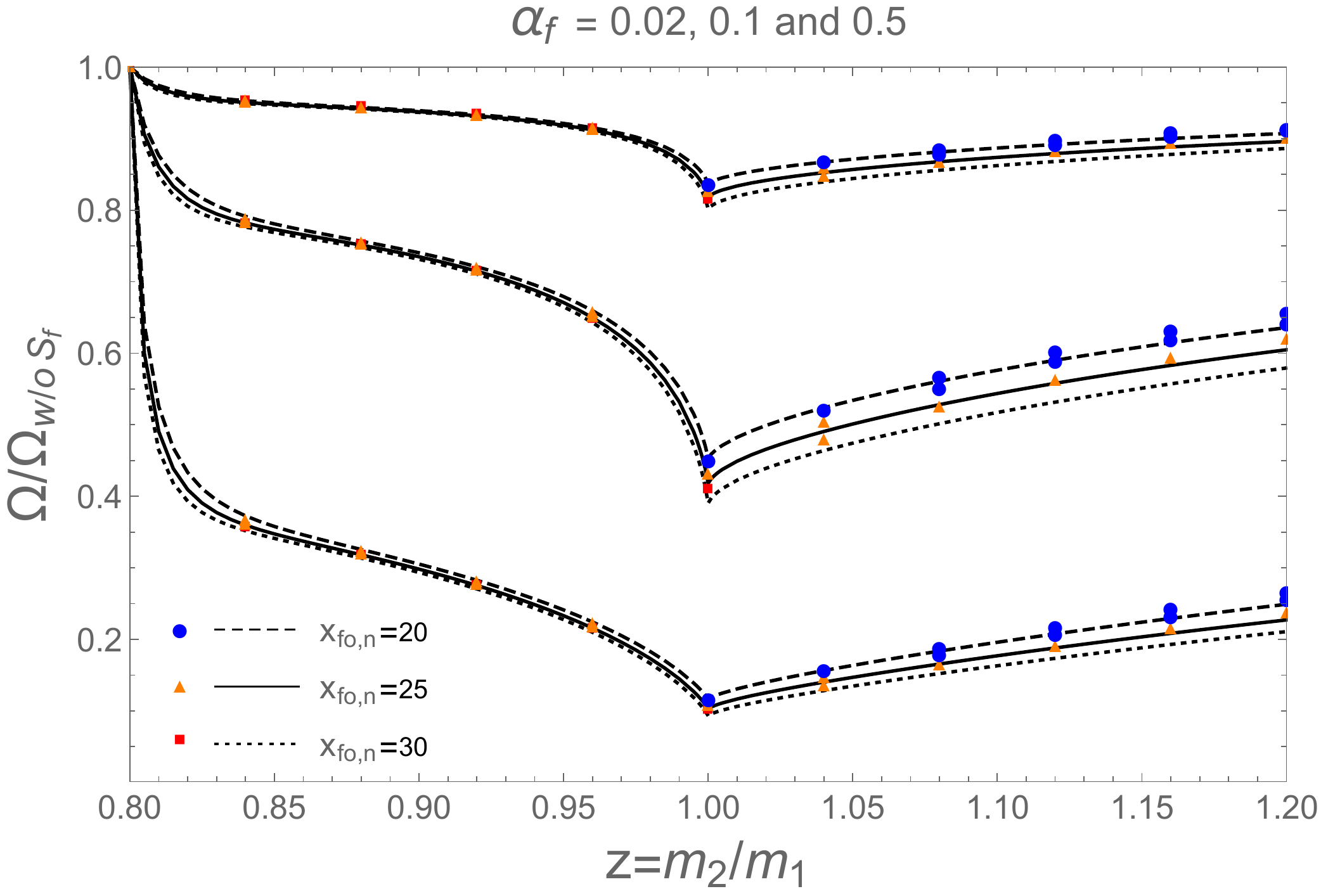} & 
\hspace{-0.3cm}
\includegraphics[height=5.8cm]{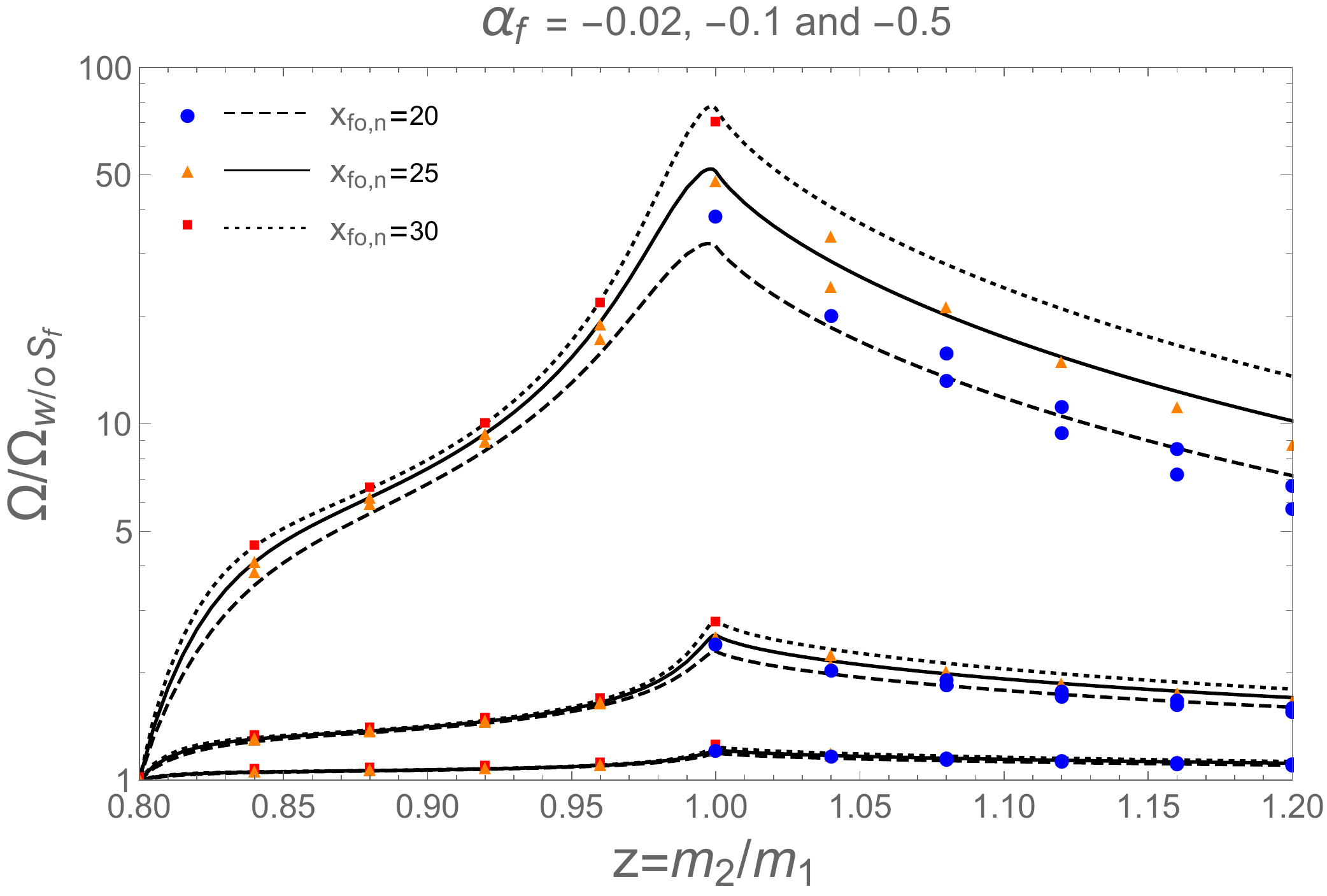} \\
\end{tabular}
\end{center}   
\caption{\label{fig: Omega ratio}\it
The ratio of DM relic densities with and without the FSS effect computed from Eq.~(\ref{eq: ratio}), as a function of the mass ratio of the final and initial state particles. The dashed, solid and dotted lines are for the representative freeze-out values of $x_{\rm fo,n} = 20$, $25$ and $30$, respectively, without the FSS effect.
The three groups of lines from top to bottom are for $\alpha_f = 0.02$, $0.1$ and $0.5$ in the left panel, while they are for $- \, 0.5$, $- \, 0.1$ and $- \, 0.02$ in the right panel.
Also plotted are ratios calculated directly from Eq.~(\ref{eq: single_Boltzmann}), using three sets of $(a^\prime/{\rm GeV^{-2}}, m_1/{\rm GeV}) = (10^{-9}, 10^3)$, $(10^{-9}, 10^4)$, $(10^{-7}, 10^4)$, and $g_1 = 2$. 
According to whether the freeze-out, without including the FSS effect, happens before $m_1/T = 22.5$, between $22.5$ and $27.5$, or after $27.5$, these ratios are marked with blue dots, orange triangles, or red squares, respectively.}
\end{figure}

Fig.~\ref{fig: Omega ratio} shows the estimation of the ratio. 
We choose three representative values of $x_{\rm fo,n}$, $20$, $25$ and $30$, corresponding to the dashed, solid and dotted lines, respectively.
The three groups of lines from top to bottom are for $\alpha_f = 0.02$, $0.1$ and $0.5$ in the left panel, while they are for $\alpha_f = - \, 0.5$, $- \, 0.1$ and $- \, 0.02$ in the right panel. 
Again all lines merge at $z = 0.8$, since we switch off the FSS effect at and below this value of $z$.  
From the solid lines, we can see that the FSS effect results in a more than $\sim 5 \%$ ($23 \%$, $65 \%$) reduction to the relic abundance for $\alpha_f = 0.02$ ($0.1$, $0.5$) for $z \ge 0.85$, and the reduction reaches $\sim 18 \%$ ($58 \%$, $90 \%$) at $z = 1$;
for $\alpha_f = - \, 0.02$ ($- \, 0.1$, $- \, 0.5$), the FSS effect increases the relic abundance by more than $\sim 6 \%$ ($32 \%$, a factor of $\sim 4.7$) for $z \ge 0.85$, and the value reaches $\sim 22 \%$ ($2.6$, $51$) at $z = 1$. 
Therefore, indeed the FSS effect can be important in DM relic abundance calculations, and an electroweak interaction sized $\alpha_f$ is already large enough to make a correction well beyond the percent level accuracy of the observational value. 

In order to check the estimation, we start from Eq.~(\ref{eq: single_Boltzmann}) and choose three sets of values of $(a^\prime, m_1)$ to calculate DM relic densities with and without the FSS effect, and then we get the ratios. 
The sets are $(10^{-9} \, {\rm GeV^{-2}}, 10^3 \, {\rm GeV})$, $(10^{-9} \, {\rm GeV^{-2}}, 10^4 \, {\rm GeV})$ and $(10^{-7} \, {\rm GeV^{-2}}, 10^4 \, {\rm GeV})$. 
For all sets, we use $g_1 = 2$ in the calculations. 
We compute the ratios for $\alpha_f = \pm \, (0.02, 0.1, 0.5)$, and from $z = 0.8$ to $1.2$ with an interval of $0.04$. 
In Fig.~\ref{fig: Omega ratio}, we mark each ratio according to the freeze-out value of $x$ without including the FSS effect, given by when $Y_1 / Y_{1, eq} = \sqrt 2$, as used in~\cite{Griest:1990kh}.
A ratio is marked with a blue dot if this freeze-out value of $x$ is less than 22.5, an orange triangle if it is between $22.5$ and $27.5$, and a red square if it is larger than 27.5. 
We expect that the marks should fall closely to the lines computed by using the same $\alpha_f$. 
We also expect that for each $\alpha_f$, the blue dots, orange triangles and red squares should fall closer to the dashed, solid and dotted line, respectively.
We can see that it is indeed the case, and this confirms our estimation formula Eq.~(\ref{eq: ratio}). 

\subsection{Correction to the FSS effect induced by final state particle decays}
\label{subsec: velocity cut-off}
Unlike the initial state DM particles which are stable or very long lived, the final state particles may decay. 
Consequently, we need to consider whether there is enough time for the FSS effect to happen before final state particles decay.
The former time scale is determined by the typical long-range interaction time between the two final state particles, $\sim (m_2 v_2^2)^{-1}$, while the latter is the inverse decay rate $\Gamma_2^{-1}$~\cite{Fadin:1993kg, Bardin:1993mc, Fadin:1993kt, hep-ph/9501214}. 
Then for $v_2 \lesssim \sqrt{\Gamma_2/m_2}$, the FSS effect is ineffective~\footnote{An equivalent explanation is as follows. The characteristic distance of the long-range interaction is given as the inverse of the relative momentum, $\left[(m_2/2)(2 v_2)\right]^{-1} = (m_2 v_2)^{-1}$, where $m_2/2$ is the reduced mass of the final state particles, and $2 v_2$ is their relative velocity. The typical spatial separation before the decay of any of the two final state particles is $(2 v_2)/(2 \Gamma_2) = v_2/\Gamma_2$. The Sommerfeld effect is suppressed when the former distance is larger than the latter one.}. 
The size of $\sqrt{\Gamma_2/m_2}$ is model dependent; therefore, we choose several cut-off velocities $v_{2 \, \rm cut}$ to look at their impacts on the FSS effect.
That is, in Eq.~(\ref{eq: thermal S_f v_2 int_t}) we substitute $S_f$ by $\left[(S_f - 1) H(0.6 - v_2) H(v_2 - v_{2 \, \rm cut})+ 1\right]$ in the following calculations~\footnote{By using the Heaviside step function $H(v_2 - v_{2 \, \rm cut})$, the FSS factor becomes 1 for $v_2 < v_{2 \, \rm cut}$. However, we note that the suppression of the FSS factor from $v_2 > v_{2 \, \rm cut}$ to $v_2 < v_{2 \, \rm cut}$ is a smooth transition. Nevertheless, a sharp cut-off is sufficient for an estimation of the effect of final state particle decays on the FSS factor. Similarly, in Sec.~\ref{sec:iss} we also use a sharp velocity cut-off for an estimation of the effect of the coannihilator decays on the ISS factor.}.  

\begin{figure}
\begin{center}
\begin{tabular}{c c}
\hspace{-0.3cm}
\includegraphics[height=5.8cm]{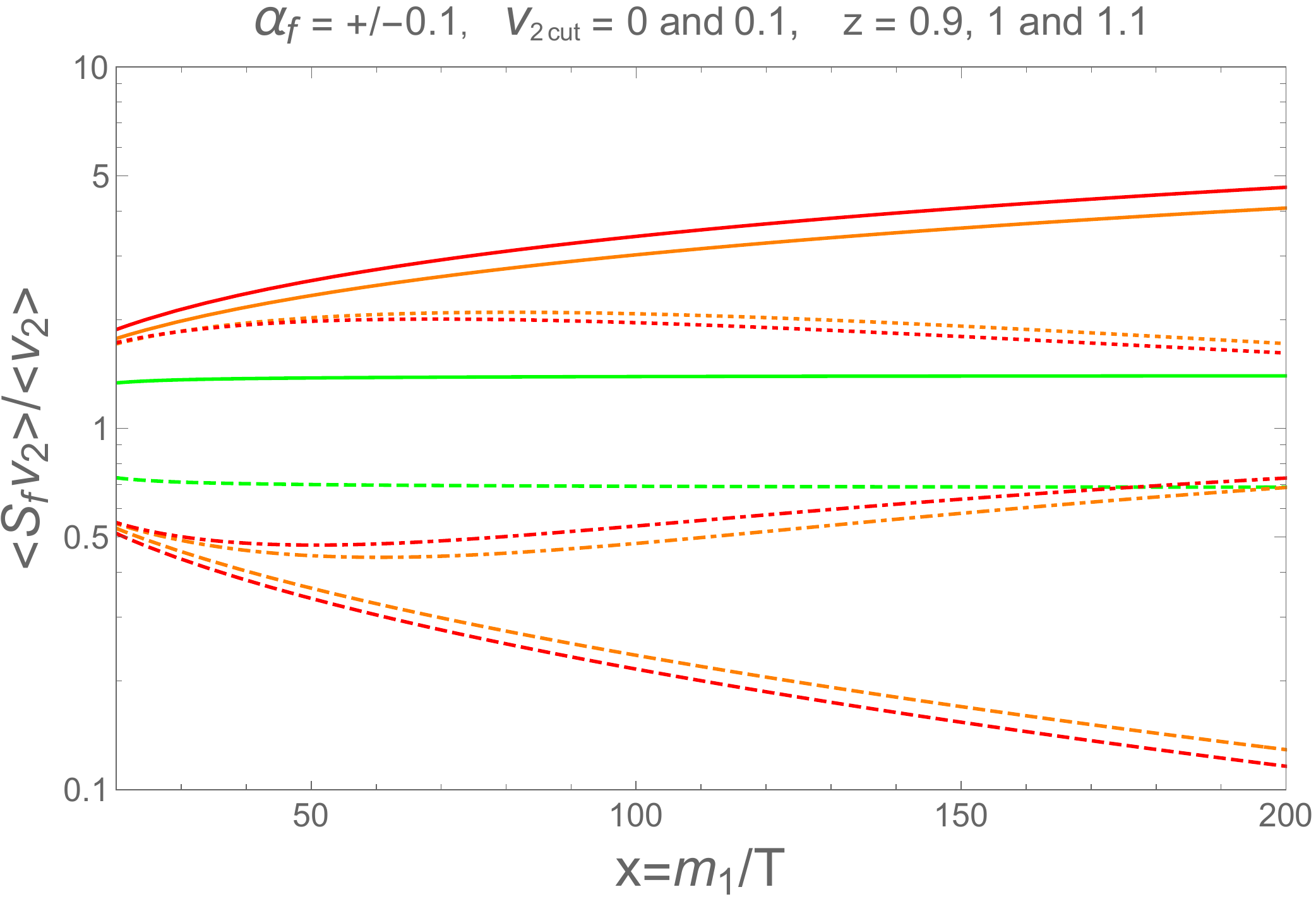} & 
\hspace{-0.3cm}
\includegraphics[height=5.8cm]{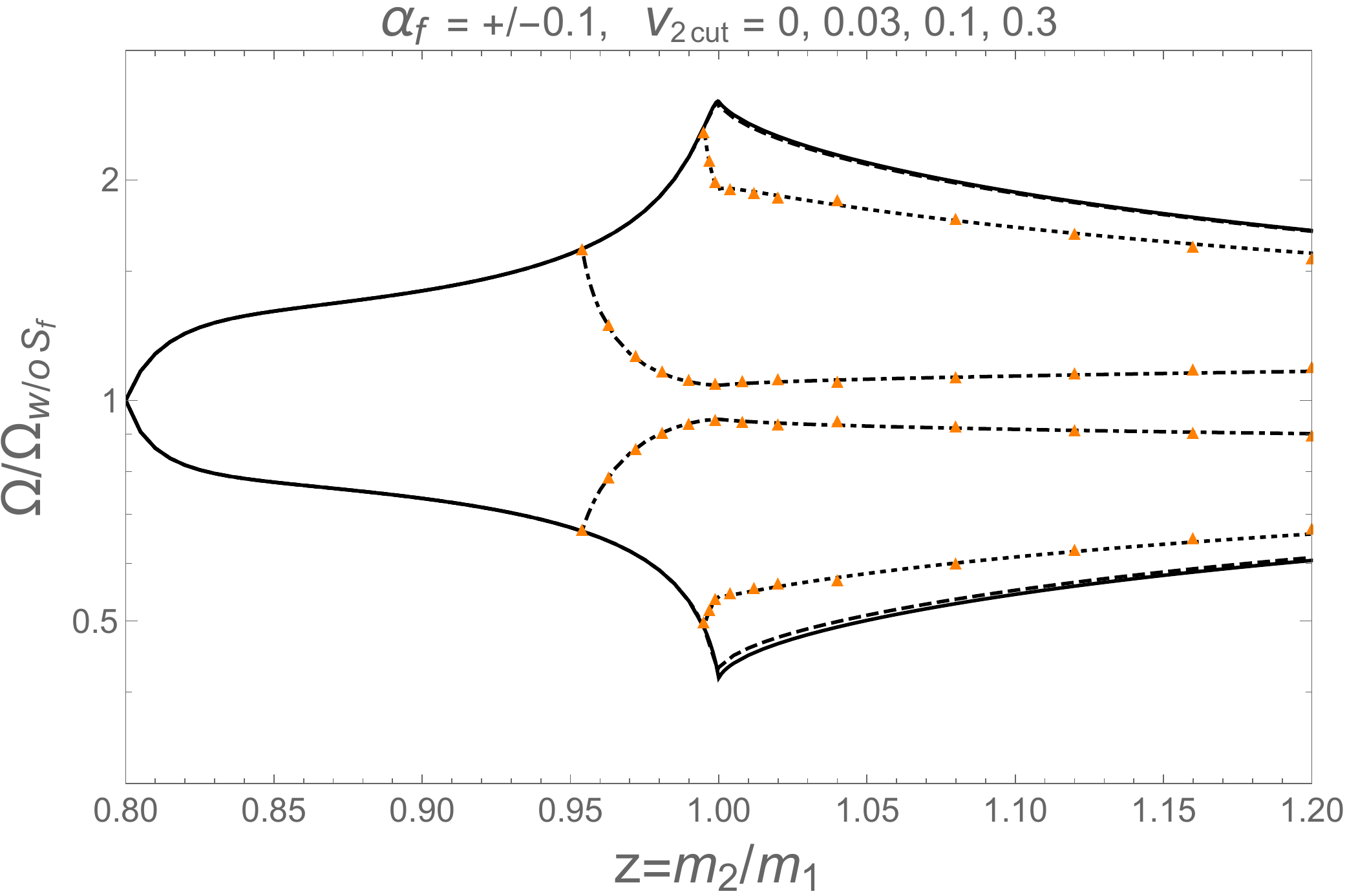} \\
\end{tabular}
\end{center}   
\caption{\label{fig: v2cut}\it
Left panel: the ratio of the thermally averaged cross sections with and without the FSS effect as a function of $x = m_1/T$, for $z = 0.9$ (green lines), $1$ (orange lines) and $1.1$ (red lines). The lines above (below) $\langle S_f v_2 \rangle / \langle v_2 \rangle = 1$ are for $\alpha_f = 0.1$ $(- \, 0.1)$. The solid and dashed lines are for $v_{2 \, \rm cut} = 0$, while the dotted and dash-dotted lines are for $v_{2 \, \rm cut} = 0.1$. 
Right panel: the ratio of DM relic densities with and without the FSS effect computed from Eq.~(\ref{eq: ratio}), as a function of $z = m_2/m_1$. The solid, dashed, dotted and dash-dotted lines are for $v_{2 \, \rm cut} = 0$, $0.03$, $0.1$ and $0.3$, respectively. The lines above (below) $\Omega / \Omega_{\rm w/o \, S_f} = 1$ are for $\alpha_f = - \, 0.1$ $(0.1)$. In all lines $x_{\rm fo,n} = 25$ is used. 
The orange triangles are ratios calculated directly from Eq.~(\ref{eq: single_Boltzmann}) by imposing $v_{2 \, \rm cut} = 0.1$ (for marks located near the dotted lines) or $0.3$ (for marks located near the dash-dotted lines), using $(a^\prime/{\rm GeV^{-2}}, m_1/{\rm GeV}) = (10^{-7}, 10^4)$ for $1.04 \le z \le 1.2$, and $(10^{-9}, 10^4)$ for the rest. 
}
\end{figure} 

We show in the left panel of Fig.~\ref{fig: v2cut} the final state particle decay effect on the evolution of $\langle S_f v_2 \rangle / \langle v_2 \rangle$ with $x$. 
Since the behaviors of this quantity are different on the two sides of $z = 1$, as shown in Fig.~\ref{fig: sigmav_vs_z}, we study its evolution for three $z$ values, $0.9$, $1$ and $1.1$, represented by the green, orange and red lines, respectively. 
The solid (dashed) lines above (below) $\langle S_f v_2 \rangle / \langle v_2 \rangle = 1$ are for $\alpha_f = 0.1$ $(- \, 0.1)$; these lines are all for $v_{2 \, \rm cut} = 0$, that is, the final state particle decay effect is not included. 
The two dotted lines and the two dash-dotted lines are for $(\alpha_f = 0.1, v_{2 \, \rm cut} = 0.1)$ and $(\alpha_f = - \,0.1, v_{2 \, \rm cut} = 0.1)$, respectively. 
We can see that the deviations of the dotted (dash-dotted) lines from the solid (dashed) lines grow with the increase of $x$. 
This can be understood from the fact that typical velocities of final state particles decrease with the increase of $x$. 
For a given $\alpha_f$, the FSS factor $S_f$ deviates from 1 more significantly for smaller $v_2$, but particles with $v_2 < v_{2 \, \rm cut}$ do not contribute.
These two opposite $v_2$ effects compete, and the deviations of dotted and dash-dotted lines from $\langle S_f v_2 \rangle / \langle v_2 \rangle = 1$ cease to grow and then reduce around $x \sim 60$. 
We note that only the cases $z = 1$ and $1.1$ have the $v_{2 \, \rm cut} = 0.1$ lines. For the case $z = 0.9$ the minimum $v_2$ is larger than 0.1 so that $H(v_2 - v_{2 \, \rm cut})$ always equals 1 for $v_{2 \, \rm cut} = 0.1$. 

In the right panel of Fig.~\ref{fig: v2cut} we show the impact of final state particle decays on the ratio of DM relic densities with and without the FSS effect.
The lines above and below $\Omega / \Omega_{\rm w/o \, S_f} = 1$ are for $\alpha_f = - \, 0.1$ and $0.1$, respectively. 
$x_{\rm fo,n} = 25$ is used for all lines.
The solid lines are for $v_{2 \, \rm cut} = 0$, and they are the same as the middle solid lines in the two panels of Fig.~\ref{fig: Omega ratio}. 
The dashed, dotted and dash-dotted lines are for $v_{2 \, \rm cut} = 0.03$, $0.1$ and $0.3$, respectively. 
For $z < 0.995$, the dotted and solid lines completely overlap, since the minimum value of $v_2$ is larger than 0.1.
For the same reason, the dash-dotted and solid lines completely overlap for $z < 0.954$. 
We can see that the deviations of the dotted and dash-dotted lines from the solid lines quickly grow until $z = 1$, and then gradually reduce for larger $z$. 
At $z = 1$, for $\alpha_f = 0.1$, compared to the $\sim 58 \%$ reduction of the relic abundance when not considering decays of the final state particles, the values are $\sim 46 \%$ and only $6 \%$ for $v_{2 \, \rm cut} = 0.1$ and $0.3$, respectively; for $\alpha_f = - \, 0.1$, the FSS effect increases the relic abundance by a factor of $\sim 1.9$ and only $5 \%$ for $v_{2 \, \rm cut} = 0.1$ and $0.3$, respectively, compared to a factor of $\sim 2.6$ increase for $v_{2 \, \rm cut} = 0$. 
On the other hand, the deviations of the dashed lines from the solid lines are quite small. 

Similar to what we did in Fig.~\ref{fig: Omega ratio}, we check the results by calculating the FSS-corrected DM densities from Eq.~(\ref{eq: single_Boltzmann}), imposing $v_{2 \, \rm cut} = 0.1$ or $0.3$. 
We use $(a^\prime = 10^{-7} \, {\rm GeV^{-2}}, m_1 = 10^4 \, {\rm GeV})$ for the orange triangles in the range $1.04 \le z \le 1.2$, and $(a^\prime = 10^{-9} \, {\rm GeV^{-2}}, m_1 = 10^4 \, {\rm GeV})$ for the rest, because for each of these points the freeze-out value of $x$ is between 22.5 and 27.5 when the FSS factor is not included. 
All the orange triangles match well the corresponding lines, and this confirms the results. 

We conclude that the modification of the FSS effect induced by final state particle decays is 
significant when $\Gamma_2/m_2 \gtrsim \mathcal{O}(10^{-1})$, and it is already 
not negligible when $\Gamma_2/m_2 \sim \mathcal{O}(10^{-2})$. 

\section{Modification of the initial state Sommerfeld effect in DM coannihilations induced by the coannihilators' instability}
\label{sec:iss}
As a corollary, we investigate in DM coannihilation scenarios whether there are also noticeable corrections to the Sommerfeld effect between coannihilators induced by their instability. 
We consider the usual situation that the coannihilators and the DM share the same discrete symmetry which makes the DM stable (e.g., the $R$-parity in supersymmetric models~\cite{Jungman:1995df} and the $KK$-parity in Universal Extra Dimension models~\cite{Servant:2002aq}), so that a coannihilator can convert to the DM particle or other species of coannihilators through decays or scatterings. 
In order to have the two-body wave function of a pair of coannihilators modified from the plane wave, the two particles need to come together from an initial separation large enough compared to the characteristic distance of the long-range interaction, $\sim (m_c v_c)^{-1}$, where $m_c$ is the coannihilator's mass and $v_c$ is one of the coannihilators' velocity in their center-of-mass frame.   
The initial separation is give as $\sim v_c / (\Gamma_{\rm decay} + \Gamma_{\rm scatt})$, where $\Gamma_{\rm decay}$ and $\Gamma_{\rm scatt}$ are the decay and scattering rates.
If the initial separation is larger, a coannihilator would have decayed or converted away before it meets the other one~\footnote{If prefer, one can use the same argument as used in the FSS effect case by considering a time-reversal.}. 

Both the decay and scattering rates are model dependent~\footnote{In particular, the coannihilator and the DM may not have a direct coupling, but are indirectly connected by some heavy particles. For example, in the neutralino-gluino coannihilation scenario in supersymmetric models, the neutralino and the gluino are indirectly connected through a squark, and the rates can be very small so that the coannihilation may not happen, if the squark mass is very heavy~\cite{1503.07142, 1504.00504}.}. 
In this work, we consider the simplest situation by assuming that there is only one species of coannihilator, and that there is a direct coupling among a coannihilator, a DM and a massless particle, so that at the lowest order the rates can be written as $\Gamma_{\rm decay} \sim c_{\rm decay} \Delta m$ and $\Gamma_{\rm scatt} \sim c_{\rm scatt} T^3/m_1^2$, where $\Delta m$ is the mass difference between the coannihilator and the DM, $\Delta m \equiv m_c - m_1$, $c_{\rm scatt}$ and $c_{\rm decay}$ are massless parameters depending on the couplings and mixings in a specific BSM model. 
Since $c_{\rm scatt}$ and $c_{\rm decay}$ are controlled by the same interaction vertex, they usually do not differ by orders of magnitude. Then $\Gamma_{\rm scatt}$ is typically smaller than $\Gamma_{\rm decay}$ during and after the freeze-out, unless the DM and the coannihilator are very degenerate in mass. 
We will only focus on $\Gamma_{\rm decay}$ in this work.

\begin{figure}
\begin{center}
\begin{tabular}{c c}
\hspace{-0.3cm}
\includegraphics[height=5.8cm]{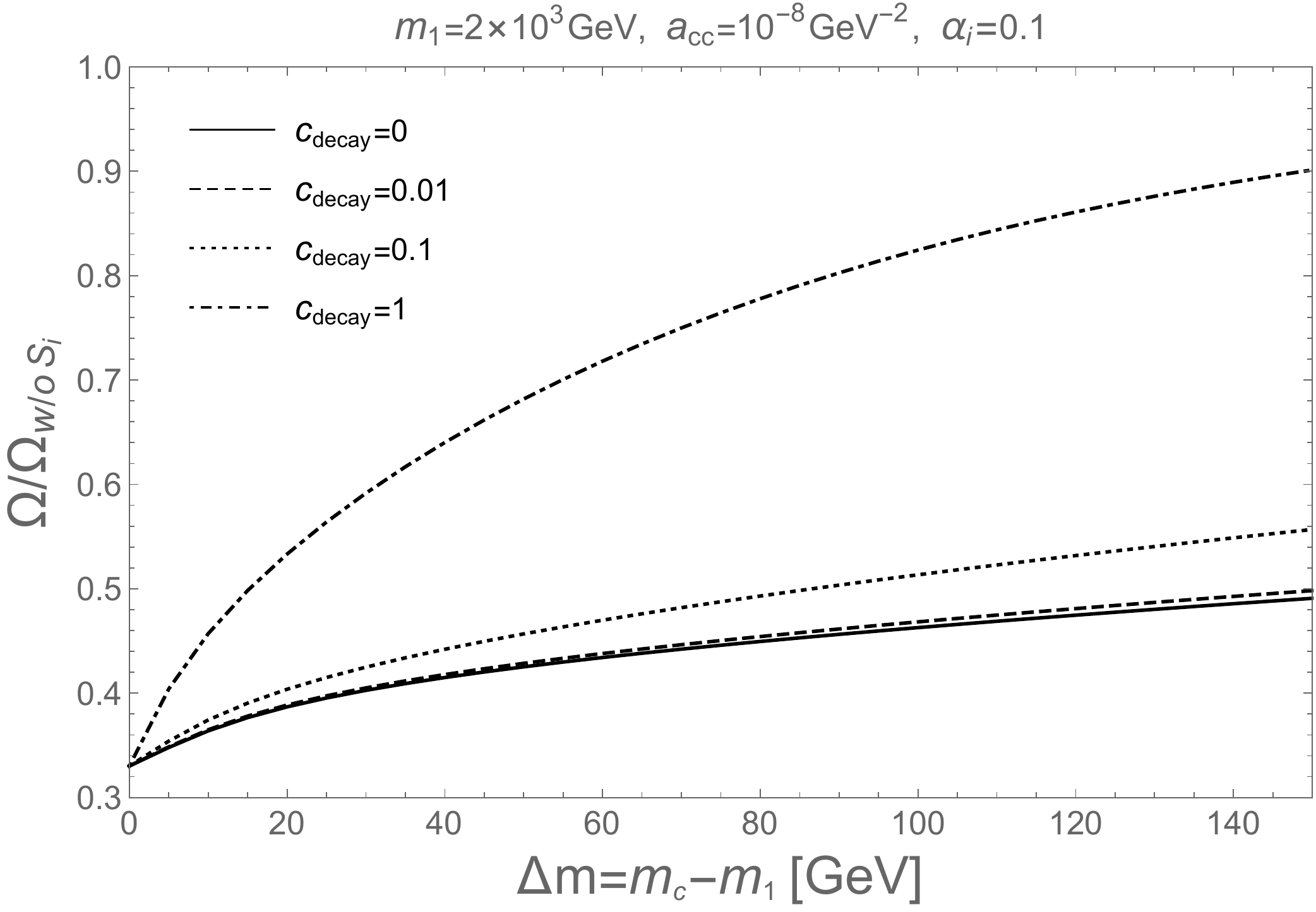} & 
\hspace{-0.3cm}
\includegraphics[height=5.8cm]{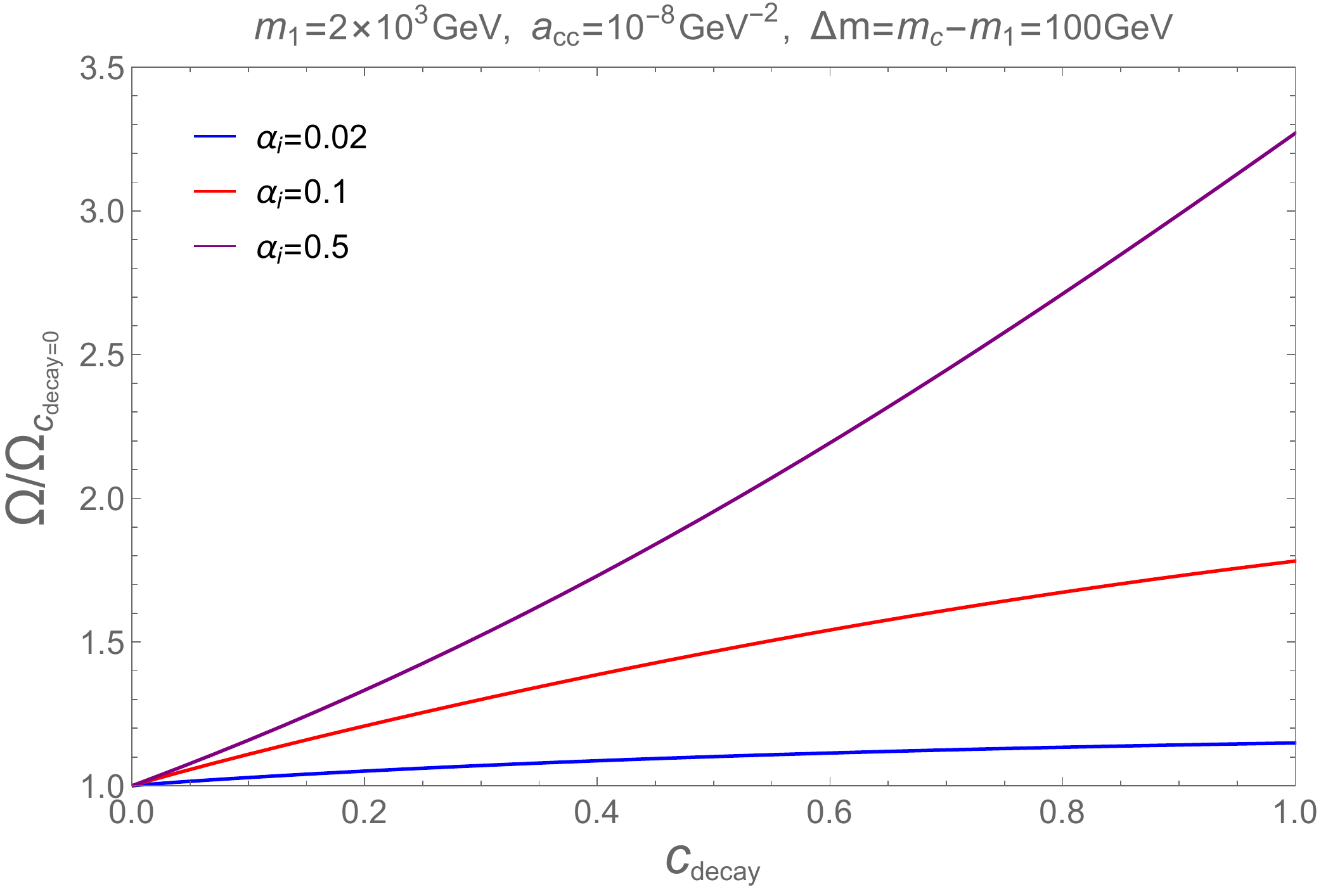} \\
\end{tabular}
\end{center}   
\caption{\label{fig: ISS plot}\it
Left panel: the ratio of DM relic densities with and without the ISS effect, as a function of the mass difference between the coannihilator and the DM particle, for $\alpha_i = 0.1$. 
The solid, dashed, dotted and dash-dotted lines are for $c_{\rm decay} = 0$, $0.01$, $0.1$ and $1$, respectively. 
Right panel: the ratio of DM relic densities with and without considering the decay of the coannihilator, as a function of the decay parameter $c_{\rm decay}$, for $\Delta m = 100 \, {\rm GeV}$. 
The blue, red and purple lines are for $\alpha_i = 0.02$, $0.1$ and $0.5$, respectively. 
For both panels, we use $g_1 = g_c = 2$, $m_1 = 2 \times 10^3 \, {\rm GeV}$ and $a_{cc} = 10^{-8} \, {\rm GeV^{-2}}$. 
}
\end{figure}

Similar to the FSS effect case, we estimate the impact of the instability of the coannihilators on the ISS effect by applying a velocity cut-off on $v_c$, given as $v_{c \rm \, cut} = \sqrt{c_{\rm decay} \Delta m / (m_1 + \Delta m)}$, below which we switch off the Sommerfeld factor between two coannihilators. 
We consider the $s$-wave annihilation cross section between a pair of coannihilators, and assume that there is a Coulomb-like potential between them, given as $V(r) = - \alpha_i / r$.
Similar to Eqs.~(\ref{eq: S_f}) and (\ref{eq: thermal S_f v_2}), the ISS factor is 
\beq
S_i = \frac{\pi \alpha_i / v_c}{1 - e^{-\pi \alpha_i / v_c}} \, , 
\label{eq: S_i}
\eeq
and the thermally averaged cross section is 
\beq
a_{cc} \langle S_i \rangle 
= a_{cc} \int_0^{\infty} 2 \left(\frac{m_c/2}{2 \pi T}\right)^{3/2} 4 \pi (2 v_c)^2 e^{- \frac{(2 v_c)^2 m_c/2}{2 T}} \left[(S_i - 1) H(v_c - v_{c \rm \, cut}) + 1\right]\, dv_c \, ,
\label{eq: thermal S_i}
\eeq
where $a_{cc}$ is the $s$-wave cross section without considering the ISS factor. 
Note that we do not need to write the term in the square bracket as $\left[(S_i - 1) H(0.6 - v_c) H(v_c - v_{c \rm \, cut}) + 1\right]$, as we did for the FSS case, because during the freeze-out and later the incoming coannihilators with relativistic velocities are in the tail of the Maxwell-Boltzmann velocity distribution~\footnote{We have explicitly checked that the curves in Fig.~\ref{fig: ISS plot} do not change  if we impose the $H(0.6 - v_c)$ factor.}.

For simplicity, we assume that for the parameter space we will consider, the effective annihilation cross section $\langle\sigma v \rangle_{eff}$ is always dominated by the above cross section, and we neglect the contributions from the DM-DM and DM-coannihilator (co)annihilation cross sections. That is, we use 
\beq
\langle\sigma v \rangle_{eff} = a_{cc} \langle S_i \rangle \frac{g_c^2 (1+\Delta m/m_1)^3 e^{-2 x \Delta m / m_1}}{g_{eff}^2} \, ,
\label{eq: sigma_eff}
\eeq
where $g_c$ is the degrees of freedom of coannihilators. $g_{eff}$ is the effective degrees of freedom, given as 
\beq
g_{eff} \equiv g_1 + g_c (1+\Delta m/m_1)^{3/2} e^{- x \Delta m / m_1} \, .
\eeq
Also, if the coannihilator is not its own antiparticle, then $a_{cc}$ should be understood as $a_{c\bar{c}}$, and we neglect the particle-particle and antiparticle-antiparticle cross sections in that case. 

The DM relic abundance can be solved by substituting in Eq.~(\ref{eq: single_Boltzmann}) $\langle \sigma  v\rangle$ by $\langle \sigma  v\rangle_{eff}$, $Y_1$ by $Y \equiv Y_1 + Y_c$ and $Y_{1,eq}$ by $Y_{eq} \equiv Y_{1,eq} + Y_{c,eq}$, where $Y_{c,eq} = n_{c,eq}/s = \frac{T}{2 \pi^2} g_c m_c^2 K_2 (m_c/T) / s$; then, in Eq.~(\ref{eq: relic abundance}) substitute $Y_{1,0}$ by $Y_{0}$.

Fig.~\ref{fig: ISS plot} shows the impact of the coannihilators' instability on DM relic abundance.
As an example, we choose $g_1 = g_c = 2$, $m_1 = 2 \times 10^3 \, {\rm GeV}$ and $a_{cc} = 10^{-8} \, {\rm GeV^{-2}}$. 

In the left panel, we fix $\alpha_i = 0.1$, and plot the ratio of DM relic densities with and without the ISS effect, as a function of $\Delta m$, for several choices of $c_{\rm decay}$. 
$\Omega_{\rm w/o \, S_i} h^2$ is calculated by letting $\langle S_i \rangle = 1$ in Eq.~(\ref{eq: sigma_eff}), and it is about $0.096$ and $1.2$ at $\Delta m = 0$ and $100 \, {\rm GeV}$, respectively.  
The solid line is the ratio when not considering the instability of the coannihilator, namely, $c_{\rm decay} = 0$ so that $v_{c \rm \, cut} = 0$.  
The dashed, dotted and dash-dotted lines are for $c_{\rm decay} = 0.01$, $0.1$ and $1$, respectively. 
The deviations of these lines from the solid line are getting larger with the increase of $\Delta m$, which controls the decay rate of the coannihilator for a given $c_{\rm decay}$.
The impact of the instability on the ISS effect is small for $c_{\rm decay} = 0.01$, while it becomes visible for $c_{\rm decay} = 0.1$ or larger. 
For instance, without considering the instability of the coannihilator, at $\Delta m = 40 \, {\rm GeV}$ the ISS effect can reduce $59 \%$ of the DM relic abundance, while the reduction becomes $56 \%$ for $c_{\rm decay} = 0.1$  and $36 \%$ for $c_{\rm decay} = 1$.   
At $\Delta m = 100 \, {\rm GeV}$, compared to the $54 \%$ reduction when taking $c_{\rm decay} = 0$, the number becomes only $18 \%$ if $c_{\rm decay} = 1$. 

In the right panel, we fix $\Delta m = 100 \, {\rm GeV}$, and plot the ratio of DM relic densities with and without considering the instability of the coannihilator, as a function of $c_{\rm decay}$, for several choices of $\alpha_i$. 
The blue, red and purple lines are for $\alpha_i = 0.02$, $0.1$ and $0.5$, respectively. 
All lines tilt up because the strength of the ISS effect is more suppressed when the coannihilators became increasingly unstable. 
The effect of the instability is more significant for larger $\alpha_i$.
For example, compared to a stable coannihilator, an unstable coannihilator with $c_{\rm decay} = 0.4$ can make the DM relic abundance larger by $9 \%$, $39 \%$ and $73 \%$ for $\alpha_i = 0.02$, $0.1$ and $0.5$, respectively. 

We conclude that when the DM and the coannihilator are not very degenerate in mass, for $\Gamma_{\rm decay} / \Delta m \gtrsim \mathcal{O}(10^{-1})$, the modification of the ISS effect induced by the decay of the coannihilator may need to be considered for an accurate calculation of the DM relic abundance. Such modification can be quite significant when the long-range force between the coannihilators is of the strong interaction or stronger size for $\Delta m / m_1 \gtrsim 1/20$. 

\section{Summary}
\label{sec:sum}
We have studied in this paper the final state Sommerfeld effect on DM relic abundance. 
This effect occurs when the DM annihilation products move non-relativistically and there is some long-range force between them, so that the wave function of the final state particles is modified from the plane wave.  
As a proof of concept, we consider the case that two WIMP DM particles annihilate into two equal mass particles, and calculate the thermally averaged $s$-wave FSS factor arising from a Coulomb-like potential between the two final state particles. 
We show the dependence of the FSS effect on the strength of the long-range interaction, as well as on the mass ratio of the final and initial state particles. 
We find that the impact of the FSS effect on DM relic abundance can be significant, and an electroweak sized long-range interaction is already large enough to make a correction well beyond the current percent level observational accuracy.

While the physical origin of the FSS effect is similar to the well-studied initial state Sommerfeld effect between two stable annihilating DM particles, an additional point of the former is that the final state particles are unstable, so that the FSS effect may not have enough time to happen before the particles decay. 
We find that when the mass ratio of the final and initial state particles is close to 1 and larger, the decay can suppress non-negligibly the FSS effect if the decay rate is larger than $1 \%$ of the final state particle's mass, and if larger than $10 \%$ the suppression becomes significant. 

As a corollary to the above point, we study the impact of the instability of the coannihilators on the initial state Sommerfeld effect in the calculations of the DM relic abundance in the coannihilation scenario. 
Here the instability comes from decays and scatterings of a coannihilator into the DM and other species of coannihilators. 
Since the decay rate usually dominates over the scattering rate unless the DM and the coannihilator are very degenerate in mass, we focus on the decay rate in this work.  
We find that the decay of the coannihilator makes a non-negligible correction to the DM relic abundance when the decay rate is more than $10 \%$ of the mass difference between the coannihilator and the DM particle. 
If the long-range interaction between the coannihilators is of the strong interaction size or more, the correction can be quite large when the mass difference is over $\sim 1/20$ of the DM mass. 

Before we close, we note that other types of long-range force, for example a Yukawa potential, can also give rise to the FSS effect when the final state particles move non-relativistically. 
Since we consider two equal mass final state particles in this work, the condition of non-relativistic moving is satisfied when the final state particle's mass is close to the initial state one. However, for the situation that the masses of the two final state particles are different, or for the final states in for instance 2-to-3 annihilation, the parameter space of at least two final state particles being non-relativistic will be different. 
Also, besides its effect in the calculations of the DM relic abundance in the early Universe, the FSS effect may play a role in the indirect searches for DM in the late Universe. 
Finally, for the impact of the instability of the coannihilators on the ISS effect, while we focus on the decay of the coannihilators in this work, the scattering is expected to be the dominant way of the coannihilator $\leftrightarrow$ DM conversion when the coannihilator and the DM are very degenerate in mass. This scenario may be worth to be explored, since it has interesting collider signals (see e.g.~\cite{Citron:2012fg, Desai:2014uha, 1504.00504, Nagata:2017gci, Abdughani:2019wss}) and it may help answer the question of how heavy the DM can be in the WIMP DM coannihilation scenarios~\cite{1503.07142, 1812.02066}. 

\section*{Acknowledgments}
The authors thank the hospitality of the University of G$\ddot{\rm o}$ttingen and Kavli IPMU where part of this work was carried out. 
X.C. is partially supported by NSFC grant 11801588 and by Guangdong Natural Science Foundation grant 2018A030313273. 
F.L. is supported by the One Hundred Talent Program of Sun Yat-sen University, China. 


\bibliographystyle{aps}
\bibliography{ref}

\end{document}